\documentclass[prd,amsmath,aps,twocolumn]{revtex4}

\usepackage{epsfig,hyperref}
\usepackage{amssymb}
\usepackage{amsfonts}

\newcommand{\ts}{{\theta}}
\newcommand{\tb}{{\overline \theta}}
\newcommand{\cb}{{\overline c}}

\begin{document}

\author{Matthieu Tissier}
\email{tissier@lptmc.jussieu.fr}
\affiliation{Laboratoire de Physique Th\'eorique de la Mati\`ere Condens\'ee,
Universit\'e Pierre et Marie Curie, 4 Place Jussieu 75252 Paris CEDEX 05, France}
\author{Nicol\'as Wschebor}
\email{nicws@fing.edu.uy}
\affiliation{Instituto de F\'{\i}sica, Facultad de Ingenier\'{\i}a,Universidad de la Rep\'ublica,J.H.y Reissig 565, 11000
Montevideo, Uruguay}
\title{Gauged supersymmetries in Yang-Mills theory}

\vspace{0.8cm}
\begin{abstract}
In this paper we show that Yang-Mills theory in the
Curci-Ferrari-Delbourgo-Jarvis gauge admits some up to now unknown
local linear Ward identities. These identities imply some
non-renormalization theorems with practical simplifications for
perturbation theory. We show in particular that all renormalization
factors can be extracted from two-point functions. The Ward identities
are shown to be related to supergauge transformations in the
superfield formalism for Yang-Mills theory. The case of non-zero
Curci-Ferrari mass is also addressed.
\end{abstract}
\maketitle
{
}
\section{Introduction}

In Yang-Mills theories, for almost all calculations aside from lattice
simulations of gauge-invariant quantities, one needs to fix the
gauge. In order to choose a sort of ``optimal'' gauge-fixing among a
large number of possibilities, one would like to preserve as many
properties of the non-fixed gauge theory as possible. In particular,
it is convenient to choose a gauge fixing that preserves Lorentz
invariance, the global color symmetry group, the renormalizability of
the theory, its locality and BRST symmetry
\cite{Becchi74,Becchi75,Tyutin75} seen as a non trivial subgroup of
the gauge symmetry. Of course, one also wants the resulting model to
be physically acceptable preserving, in particular, the
unitarity. There exist gauge-fixings that, at the perturbative level,
satisfy all these requirements. The most popular are the linear
covariant gauges, including in particular the Landau gauge. However,
such gauge-fixings are ambiguous because of the Gribov copies problem
\cite{Gribov77,Singer78,Sharpe84}. One manifestation of this problem
is that if one tries to construct a nonperturbative version of the
BRST symmetry on the lattice, the expectation value of gauge-invariant
quantities is an undefined $0/0$ expression. This is sometimes called
the Neuberger's 0 problem \cite{Neuberger86,Neuberger86b}. These
zeroes originate from the compensation in the functional integral of
the contributions of pairs of Gribov copies that come with opposite
weights.  One therefore faces the alternative of either working with a
gauge-fixing with a Gribov ambiguity or loose one of the above
mentioned properties. In fact, recent works propose a third option, that is to calculate some gauge-invariant quantities without fixing the gauge (see \cite{Arnone07} and references therein).

If one chooses the second option, one can, for example:
\begin{itemize}
\item Use the axial gauge that explicitly breaks Lorentz invariance
  but does not have Gribov problem.
\item Use the Maximal Abelian gauge that breaks the global color symmetry
  group; in this case, the partial gauge fixing to the maximal abelian
  subgroup of the gauge group has been proven to avoid the Gribov
  problem \cite{Schaden98}.
\item Use the absolute Landau gauge by imposing a global extremization
  condition of a certain functional (see, for example,
  \cite{vanBaal97}); however no local action is known to implement
  this gauge-fixing and the very useful BRST symmetry is also
  lost. Moreover, no efficient algorithm is known to implement that
  idea in practice.
\end{itemize}

In this paper we will follow a more heterodox strategy, which consists
in taking the Curci-Ferrari (CF) model \cite{Curci76,Curci76b} that
corresponds to the Yang-Mills theory in a particular gauge,
supplemented with a mass for gluons and ghosts. This model is not
unitary \cite{Curci76b,Ojima81,deBoer95} but the presence of the
masses lifts the degeneracy of contributions coming from different
Gribov copies and therefore regularizes the Neuberger's zero
\cite{Kalloniatis05,Ghiotti06}. If one studies the model directly at
zero mass, one has a standard gauge fixing sometimes called
Curci-Ferrari-Delbourgo-Jarvis (CFDJ) gauge
\cite{Curci76,Curci76b,Delbourgo81}, with all good properties,
including unitarity, except that it has a Gribov ambiguity. It is
actually possible to have unitarity {\it and} regularize the 0/0
expressions by computing physical observables in the massive theory
and then taking the limit of vanishing masses. 

The mass term can also be seen as a source for the dimension-two
composite operator $\frac 12 (A_\mu^a)^2+\xi_0 \bar c^a c^a$ that
attracted a lot of attention recently in relation with nonperturbative
effects on the behavior of the correlation functions of ghosts and
gluons (see for example \cite{Gubarev00,Kondo01,Boucaud01,Gracey02}).

All these reasons strongly motivate the use of the CF model. However,
this model is not widely used in practice, mainly because it seems
much more cumbersome than the linear gauges. For instance it has a
four-ghosts interaction. In this paper we show that this widespread
prejudice should be reconsidered. We will show that beside the large
symmetry group of the CF model, there exist local transformations that
induce very simple variations of the action. Therefore, although these
transformations are not symmetries of the full action, one can deduce
from them useful linear Ward identities. We show that there are
actually underlying symmetries associated with these transformations
that clearly appear in the superspace formulation of Yang-Mills theory
\cite{Bonora80, Bonora81, Baulieu81, Delbourgo81}.  In this
formulation, gluons and ghosts are part of a single supervector in a
superspace with 4 bosonic coordinates and 2 anticonmuting Grassmann
coordinates. We show that the transformations associated with the new
Ward identities are, in fact, supergauge transformations. The
associated identities allow us to deduce non-renormalization theorems
reducing the number of independent renormalization factors from five
\cite{deBoer95} to three.  The situation is very close to the
gauge-fixed abelian theories where gauge transformations are not
symmetries of the full bare action but allow one to deduce linear Ward
identities. This, in turn, implies that gauge-fixing terms are not
renormalized.  Another question addressed in the present paper is the
meaning of the Curci-Ferrari mass in the superspace formulation. We
show that it can be seen as a consequence of a curvature in the
grassmannian sector of the superspace.  Finally we show that these
non-renormalization theorems imply practical simplifications for
perturbation theory: all the renormalization factors can be extracted
from two external legs diagrams only.

The paper is organized as follows: In section \ref{chap_model}, we
review the model and its symmetries in the massless case. We also
derive the new Ward identities. In section \ref{nrt}, we analyze the
renormalization properties of the model and deduce a
non-renormalization theorem. In section \ref{chap_mass} we generalize
the results of the two previous sections to take into account the CF
mass and deduce another non-renormalization theorem. In section
\ref{perttheory} we analyze the consequences of these results for
perturbation theory. In section \ref{superspace} we review the
superspace formulation of Yang-Mills theory and interpret in this
context the mentioned Ward identities in terms of supergauge
transformations. We also give an interpretation of the CF model in the
superspace formulation. Finally, we give our conclusions in section
\ref{conclusion}.

\section{The action and its symmetries}
\label{chap_model}

In this section, we analyze the CF model with vanishing masses,
i.e. the Yang-Mills theory in the CFDJ gauge
\cite{Curci76,Curci76b,Delbourgo81}.  We will consider here the model
in a four dimensional euclidean space without including matter, but
most of the results can be generalized to minkowskian space and the
inclusion of matter does not modify the main results. The gauge-fixed
lagrangian reads
\begin{equation}
\mathcal{L}=\mathcal{L}_{\rm{YM}}+\mathcal{L}_{\rm{GF}}.
\end{equation}
$\mathcal{L}_{\rm{YM}}$ is the Yang-Mills lagrangian:
\begin{equation}
\mathcal{L}_{\rm{YM}}=\frac{1}{4}F_{\mu\nu}^aF_{\mu\nu}^a,
\end{equation}
$F_{\mu\nu}^a=\partial_\mu A_\nu^a-\partial_\nu A_\mu^a+g_0
f^{abc}A_\mu^b A_\nu^c$ is the bare field strength, $g_0$ is the bare
gauge coupling, $A_\mu$ is the gauge field, and $f^{abc}$ denotes the
structure constants of the gauge group that are chosen completely
antisymmetric.  $\mathcal{L}_{\rm{GF}}$ is the gauge-fixing term, which
includes a ghost sector. It takes the form:
\begin{equation}
\label{cfaction}
\begin{split}
\mathcal{L}_{\rm{GF}}&=
\frac{1}{2}\partial_\mu \bar c^a (D_\mu c)^a
+\frac{1}{2}(D_\mu \bar c)^a \partial_\mu c^a+\frac{\xi_0}{2}h^ah^a \\
&+ih^a\partial_\mu A_\mu^a 
-\xi_0\frac{g_0^2}{8}(f^{abc}\bar c^bc^c)^2.
\end{split}
\end{equation}
Here, $c$ and $\bar c$ are ghost and antighosts fields respectively,
and $(D_\mu \varphi)^a= \partial_\mu \varphi^a + g_0 f^{abc} A_\mu^b
\varphi^c$ is the covariant derivative for any field $\varphi$ in the
adjoint representation. The main interest of the CFDJ lagrangian
(\ref{cfaction}) is that the ghost-antighost exchange symmetry is
explicit and that it preserves the linear realization of some
continuous symmetries \cite{Delbourgo81}. This is not the case if the
Lagrange multiplier $h^a$ is introduced, as often done, in a
non-symmetric way:
\begin{equation}
\label{jaugenonsym}
\begin{split}
\mathcal{L}_{\rm{GF}}^{\rm {ns}}&=
\partial_\mu \bar c^a (D_\mu c)^a
+\frac{\xi_0}{2}h^ah^a +ih^a \partial_\mu A_\mu^a\\
&-i\frac {\xi_0} 2 g_0 f^{abc} h^a \bar c^bc^c
-\xi_0\frac{g_0^2}{4}(f^{abc}\bar c^bc^c)^2.
\end{split}
\end{equation}
However, these two versions of the CF model are in fact equivalent:
indeed, one obtains (\ref{jaugenonsym}) by performing the change of
variables $ih^a\to ih^a+\frac{g_0}{2}f^{abc}\bar c^b c^c$ in
(\ref{cfaction}).

Note that the considered gauge-fixing lagrangian is different from the
more standard linear gauge fixing:
\begin{equation}
\label{linearaction}
\mathcal{L}_{\rm{GF}}^{\rm{linear}}=\partial_\mu \bar c^a (D_\mu c)^a+\frac{\xi_0}{2}h^ah^a+ih^a\partial_\mu A_\mu^a.
\end{equation}
One cannot obtain one from the other by a change of variables
in the fields.  However, all these gauge fixings coincide in the
particular case of the Landau gauge limit $\xi_0 \to 0$. In fact
(\ref{jaugenonsym}) and (\ref{linearaction}) are identical in this
limit.

Let us list the symmetries of the gauge-fixing lagrangian
(\ref{cfaction}): 
\begin{enumerate}
\item[a)] The euclidean symmetries of the
spacetime. 
\item[b)] The global color symmetry.
\item[c)] The already mentioned
ghost conjugation symmetry: $c^a \to \bar c^a$, $\bar c^a \to -c^a$
without modifying the other fields. This symmetry allows one to obtain
most of the relations of this paper by conjugating those explicitly
considered.
\item[d)] The continuous symplectic group $SP(2,\mathbb{R})$
\cite{Delduc89} with generators $N$, $t$ and $\bar t$ defined by:
\begin{equation}
\label{symmsp2}
\begin{array}{lll}
t A_\mu^a=0 &\hspace{.8cm}\bar t A_\mu^a= 0&\hspace{.8cm}N A_\mu^a= 0\\
t c^a= 0 &\hspace{.8cm}\bar t c^a= -\bar c^a&\hspace{.8cm}N c^a= c^a\\
t \bar c^a= c^a &\hspace{.8cm}\bar t \bar c^a= 0&\hspace{.8cm}N \bar c^a= - \bar c^a\\
t h^a=0 &\hspace{.8cm}\bar t h^a= 0&\hspace{.8cm}N h^a= 0.\\
\end{array}
\end{equation}
$N$ is associated with the ghost-number conservation. One observes
that $A$ and $h$ are singlets while $c$ and $\bar c$ form a doublet of
this group. Note that $t$ and $\bar t$ have ghost number 2 and -2
respectively.
\item[e)] The model is also invariant under the nonlinear
BRST and anti-BRST symmetries:
\begin{align}
\label{symmbrst}
&s A_\mu^a=(D_\mu c)^a\nonumber,\\ 
&\bar s A_\mu^a= (D_\mu \bar c)^a,\nonumber\\
&s c^a= -\frac{g_0}{2} f^{abc} c^bc^c,\nonumber\\
&\bar s \bar c^a =-\frac{g_0}{2} f^{abc} \bar c^b \bar c^c,
\nonumber\\
&s \bar c^a =ih^a -\frac{g_0}{2} f^{abc}\bar c^b c^c,\nonumber\\
&\bar s c^a=-ih^a-\frac{g_0}{2} f^{abc} \bar c^bc^c,\nonumber\\
&s\, i h^a=\frac{g_0}{2} f^{abc} \Big(i h^bc^c+\frac{g_0}{4} f^{cde} \bar c^bc^dc^e\Big),\nonumber\\
&\bar s\, ih^a=\frac{g_0}{2} f^{abc} \Big(ih^b \bar c^c-\frac{g_0}{4} f^{cde} c^b \bar c^d \bar c^e\Big).
\end{align}
%\begin{align}
%\label{symmbrst}
%&
%\begin{array}{ll}
%s A_\mu^a=(D_\mu c)^a, &\hspace{.2cm}\bar s A_\mu^a= (D_\mu \bar c)^a,\\
%s c^a= -\frac{g_0}{2} f^{abc} c^bc^c,
%&\hspace{.2cm}\bar s c^a=-ih^a-\frac{g_0}{2} f^{abc} \bar c^bc^c,\\
%s \bar c^a =ih^a -\frac{g_0}{2} f^{abc}\bar c^b c^c,
%&\hspace{.2cm}\bar s \bar c^a =-\frac{g_0}{2} f^{abc} \bar c^b \bar c^c,
%\end{array}\nonumber\\
%&s\, i h^a=\frac{g_0}{2} f^{abc} \Big(i h^bc^c+\frac{g_0}{4} f^{cde} \bar c^bc^dc^e\Big),\nonumber\\
%&\bar s\, ih^a=\frac{g_0}{2} f^{abc} \Big(ih^b \bar c^c-\frac{g_0}{4} f^{cde} c^b \bar c^d \bar c^e\Big).
%\end{align}
These symmetries satisfy the standard nilpotency property ($s^2=\bar
s^2=\bar s s+s\bar s=0$).
\end{enumerate}

In order to deduce Slavnov-Taylor identities for these symmetries, it
is necessary to introduce sources for the variations of the fields
under BRST and anti-BRST symmetries. Since the symmetry is nilpotent,
it is sufficient to introduce sources for $s \varphi^a$, $\bar s
\varphi^a$ and $s \bar s \varphi^a$ for $\varphi^a=A_\mu^a, c^a$ and
$\bar c^a$ \cite{Alvarez-Gaume81}. For completeness, we give here:
\begin{align}
\label{variationssb}
s \bar s A_\mu^a&=i(D_\mu h)^a+\frac {g_0}2 f^{abc}\left(\bar c^b(D_\mu c)^c-(D_\mu \bar c)^b c^c\right),
\nonumber\\
s \bar s c&=-{g_0} f^{abc}\left( ih^b c^c+\frac {g_0}4 f^{cde}\bar c^b c^d c^e\right), \nonumber \\
s \bar s \bar c&=-{g_0} f^{abc}\left(ih^b \bar c^c-\frac {g_0}4 f^{cde} c^b \bar c^d \bar c^e\right).
\end{align}
Observe that
\begin{equation}
\label{defh}
ih^a=(s \bar c^a - \bar s c^a)/2
\end{equation}
so that the variations of $h^a$ can be expressed in terms of the
variations of the other fields. Consequently we do not introduce new
sources for these variations.

We therefore consider the generating functional:
\begin{align}
\label{pathintegral}
&\exp(W[J,\chi,\bar\chi,R,K,L,\bar K,\bar L,M,\alpha,\beta,\bar \beta])\nonumber\\
&=\int \mathcal{D}(A,c,\bar c,h) \exp\int d^4x\Big(-\mathcal{L}+\mathcal{L}_{\rm{sources}}\Big),
\end{align}
where
\begin{equation}
\label{lagsources}
\begin{split}
\mathcal{L}_{\rm{sources}}=&J^a_\mu A^a_\mu+\bar\chi^a c^a+\bar c^a
\chi^a +R^a h^a \\
&+ \bar K^a_\mu s
A^a_\mu + \bar s A^a_\mu K^a_\mu+ \bar L^a s c^a\\&+ L^a \bar s \bar c^a
+M^a(s\bar c^a+\bar s c^a)/2 \\&+\alpha^a_\mu s \bar s A^a_\mu
+\bar\beta^a s \bar s c^a
+ s \bar s \bar c^a\beta^a.
\end{split}
\end{equation}
We coupled the variations of the fields to the sources so that $R$ is
a singlet and ($L^a$, $\bar L^a$, $M^a$) a triplet of the
$SP(2,\mathbb{R})$ group.  We give in Table \ref{Table} the
dimensions, ghost-numbers and ghost conjugates of the sources and
fields.
\begin{table}
\begin{center}
\begin{tabular}{|c|c|c|c|c|c|c|c|c|c|c|c|c|}
\hline 
\ \ Field/Source\ \  & $A$ & $c$ &$\bar c$& $h$ &$\alpha$&$\beta$&$\bar\beta$& $K$ &$\bar K$& $L$ &$\bar L$& $M$ \\
\hline 
Dimension            & 1   &  1  &    1   & 2 & 1 & 1 & 1 & 2  &    2   &  2  &   2    & 2\\ \hline
Ghost number         & 0   &  1  &   -1   & 0 & 0 & 1& -1& 1 &    -1   & 2  &   -2    & 0 \\ \hline
Conjugation    & $A$ &$\bar c$ &$-c$ & $h$ &$\alpha$&$\bar \beta$&$-\beta$&$\bar K$&$-K$&$\bar L$  & $L$ & $-M$ \\ 
\hline
\end{tabular}
\caption{\label{Table} \it Canonical dimension, ghost number and
  (ghost) conjugation of different fields and sources.}
\end{center}
\end{table}

Simple Ward identities can be easily derived for linearly realized
symmetries (a to d). For instance, the Ward identity associated with
the symmetry of generator $t$ is:
\begin{equation}
\begin{aligned}
\int d^4x\Big(c^a \frac{\delta \Gamma}{\delta\bar c^a }+ K_\mu^a& \frac{\delta
  \Gamma}{\delta\bar K_\mu^a }-2L^a \frac{\delta \Gamma}{\delta M^a }
\\
-&M^a \frac{\delta \Gamma}{\delta \bar L^a }+\beta^a
\frac{\delta \Gamma}{\delta \bar \beta^a }\Big)=0.
\end{aligned}
\end{equation}
As usual, the Slavnov-Taylor identity \cite{Slavnov72,Taylor71}
associated to the BRST symmetry is obtained by performing the change of
variables in the functional integral $\varphi\to \varphi +\varsigma\,
s\varphi$ for all fundamental fields $\varphi$ with a constant
grassmanian parameter $\varsigma$. One obtains:
\begin{align}
\label{ST}
\int d^4&x\Bigg\{-\frac{\delta
\Gamma}{\delta \bar K_\mu^a}\frac{\delta \Gamma}{\delta A_\mu^a }-\frac{\delta \Gamma}{\delta \bar L^a}\frac{\delta \Gamma}{\delta c^a}\nonumber\\
&+\Big(ih^a-\frac{\delta\Gamma}{\delta M^a }\Big)
\frac{\delta \Gamma}{\delta \bar c^a} 
-K_\mu^a\frac{\delta \Gamma}{\delta \alpha_\mu^a}\nonumber\\
&+L^a
	\frac{\delta \Gamma}{\delta \beta^a}+\frac 12 
\Big(-i\frac{\delta \Gamma}{\delta h^a}-M^a\Big)\frac{\delta \Gamma}{\delta \bar \beta^a}\Bigg\}=0.
\end{align}
A similar equation can be deduced for the anti-BRST symmetry. However,
we will not need it here because its information can be obtained by
exploiting the ghost conjugation (c). The physical interpretation of
(\ref{ST}) is well-known.  If one evaluates it for vanishing sources
for composite operators, it says that $\Gamma$ is invariant under
$A_\mu^a\to A_\mu^a-\varsigma\,\delta \Gamma/\delta \bar K_\mu^a$,
$c^a \to c^a -\varsigma \delta \Gamma/\delta \bar L^a$, etc. The
symmetry transformation itself acquires quantum corrections.

After this review of these well-known symmetries and their
consequences, we now come to the deduction of other Ward identities
that are linear and local. The first one is the equation of motion for
the Lagrange multiplier $h^a$. It can be obtained in the usual way by
performing an infinitesimal space-time dependent shift on the $h$
field $ih^a(x)\to ih^a(x)+\hat\lambda^a(x)$. This gives:
\begin{equation}
\label{eqmoth}
\begin{split}
\frac{\delta \Gamma}{\delta h^a}=\xi_0 h^a+i&\big[\partial_\mu
  A_\mu^a+(D_\mu \alpha_\mu)^a\\&-{g_0} f^{abc}\left(\bar\beta^b c^c+\bar
  c^c \beta ^b\right)\big].
\end{split}
\end{equation}
This equation means that terms in the effective action including the
$h$ field are not renormalized. Note that the non-symmetric lagrangian
(\ref{jaugenonsym}) contains terms that couple the $h$ field
tri-linearly which prevents one to derive a simple equation as
(\ref{eqmoth}). Such terms do not exist in lagrangians
(\ref{cfaction}) and (\ref{linearaction}) giving tractable equations
of motion for $h$. Another gauge where tractable equations for the
(abelian) Lagrange multiplier can be deduced is the Maximal Abelian gauge.

In the case of linear gauge-fixing, as well as in Maximal Abelian
gauge \cite{Fazio01,Ellwanger02}, another local and linear identity
can be deduced from the equation of motion of the anti-ghost field.
We find an analogous relation here if we shift the ghost field by a
space-dependent term $\delta \bar c^a(x)=\bar \eta^a(x)$ and
simultaneously change the Lagrange multiplier according to $\delta
ih^a(x)=\frac {g_0} 2 f^{abc}\bar\eta^b(x) c^c(x)$:
\begin{equation}
\begin{aligned}
\label{eqmotc}
&-\frac {\xi_0} 2 \frac{\delta \Gamma}{\delta \bar\beta^a}-\partial_\mu
\frac{\delta \Gamma}{\delta\bar K_\mu^a }+\frac{\delta \Gamma}{\delta
  \bar c^a}-D_\mu K_\mu^a \\
&+g_0 f^{abc} \Bigg(-\bar c^bL^c+\frac 12
c^b\Big(-i \frac{\delta \Gamma}{\delta h^c}-M^c \Big) \\
& - \frac{\delta
  \Gamma}{\delta\bar K_\mu^b } \alpha_\mu^c+ \frac{\delta
  \Gamma}{\delta\bar L^b } \bar\beta^c+\Big(ih^b-\frac{\delta
  \Gamma}{\delta M^b}\Big) \beta^c\Bigg)=0.
\end{aligned}
\end{equation}
Here, contrarily to what happens in linear gauges, we obtain a third
equation by ghost conjugation.

A fourth identity can be deduced by making the following change of
variables in the functional integral:
\begin{align}
\label{transA}
\delta A_\mu^a(x)&=(D_\mu \lambda(x))^a, \nonumber\\
\delta c^a(x)&= g_0  f^{abc} c^b(x) \lambda^c(x), \nonumber\\
\delta \bar c^a(x)&= g_0  f^{abc}\bar c^b(x) \lambda^c(x), \nonumber\\
\delta h^a(x)&= g_0  f^{abc} h^b(x)  \lambda^c(x),
\end{align}
which gives the identity:
\begin{equation}
\label{eqmotA}
\begin{split}
&\Big(D_\mu \frac{\delta \Gamma}{\delta A_\mu}\Big)^a-\partial_\mu
\frac{\delta \Gamma}{\delta \alpha_\mu^a}=g_0f^{abc}\Bigg( c^c
\frac{\delta \Gamma}{\delta c^b}+ \bar c^c \frac{\delta \Gamma}{\delta
  \bar c^b}\\
&+ K_\mu^c \frac{\delta \Gamma}{\delta K_\mu^b}
+ \bar K_\mu^c \frac{\delta \Gamma}{\delta \bar K_\mu^b}+ h^c \frac{\delta
  \Gamma}{\delta h^b}+ M^c \frac{\delta \Gamma}{\delta M^b}\\
&+ L^c
\frac{\delta \Gamma}{\delta L^b}+ \bar L ^c \frac{\delta
  \Gamma}{\delta \bar L ^b}+\alpha_\mu ^c \frac{\delta \Gamma}{\delta \alpha_\mu ^b}+
\beta^c \frac{\delta \Gamma}{\delta \beta^b}+ \bar \beta^c
\frac{\delta \Gamma}{\delta \bar \beta^b} \Bigg).
\end{split}
\end{equation}
To our knowledge, no such relation was found in the linear
gauge-fixing. 

Let us stress that these four identities (Eqs. (\ref{eqmoth}),
(\ref{eqmotc}) and its conjugate, and (\ref{eqmotA})) are not fully
independent from the Slavnov-Taylor equation (\ref{ST}). Actually, the
change of variable yielding (\ref{eqmotc}) is obtained by commuting
the shift of $h$ used to deduce (\ref{eqmoth}) and the BRST
transformation that generate (\ref{ST}). Similarly, the transformation
(\ref{transA}) is obtained by commuting the anti-BRST transformation
with the transformations that leads to (\ref{eqmotc}). Observe also
that these four equations look like Ward identities for gauged linear
(super)symmetries letting aside some non-homogeneous terms that play
the role of gauge-fixing terms. As mentioned in the introduction,
these terms behave as in abelian gauge theories where gauge-fixing
preserves its bare form under the renormalization process. This is
very different from Slavnov-Taylor identities, which are non-linear in
$\Gamma$ and therefore much harder to handle. The obtention of local,
linear Ward identities is a non-trivial result and is the heart of the
present manuscript.

The equations (\ref{eqmoth}), (\ref{eqmotc}) and (\ref{eqmotA}) are
very simple and have far reaching consequences. However, to our
surprize, they have never been addressed before in the CF model. In
the next two sections, we discuss the consequences of these relations
showing, in particular, that they induce non-trivial
non-renormalization theorems for some quantities.

\section{Non-renormalization theorem for the coupling}
\label{nrt}

The four new identities derived in the previous section have many
consequences on the form of the effective action. As a concrete
example, we analyze in this section the implications on the
renormalization properties of the model.

The perturbative renormalizability of this model has been proven by
considering five renormalization factors \cite{deBoer95}, including
the renormalization of the mass. Recently, however, one of us
\cite{Wschebor07} proved two non-renormalization theorems that reduce
the number of renormalization factors from five to three. We now prove
in this section and in the following that these non-renormalization
theorems are, in fact, a direct consequence of the new identities
discussed in the previous section.

We follow the standard procedure (see for example \cite{Weinberg96})
of considering terms that can diverge by power counting and
constraining them iteratively. In a loop expansion, suppose that all
divergences have been renormalized at order $n-1$.  Divergent terms
that appear at order $n$ in the effective action, have couplings with
positive or zero dimension.  Let us call them $\Delta
\Gamma^{(n)}_{div}$, and take an infinitesimal constant $\epsilon$.
If one calls $\Gamma_{div}^{(n)}=S+\epsilon \Delta
\Gamma^{(n)}_{div}$, then, in four dimensions, the most general form
for this functional at order $n$ that satisfies the linear symmetries
(a--d), takes the form:
\begin{align}
\label{previous1}
&\Gamma^{(n)}_{div}[A,c,\bar c,h,K,\bar K,L,\bar L,M,\alpha,\beta,\bar
  \beta]=\nonumber\\
&-\int d^4x\Big\{Z_L \Big(\bar L^a L^a
-\frac{1}{4} M^a M^a\Big)+Z_K \bar K_\mu^a K_\mu^a \nonumber \\ &+
\bar K^a_\mu \tilde s A^a_\mu + \tilde{\bar s} A^a_\mu K^a_\mu + \bar
L^a \tilde s c^a+ L^a \tilde{\bar s} \bar c^a\nonumber\\ &+M^a(\tilde
s\bar c^a+\tilde{\bar s} c^a)/2 \Big\}+\hat\Gamma[A,c,\bar c,h,\alpha,\beta,\bar
  \beta].
\end{align}
We introduced the notation $\tilde s$ and $\tilde{\bar s}$ in terms
linear in $K,\bar K,L, \bar L$ and $M$ in analogy to
(\ref{lagsources}). However, for the moment, $\tilde s A_\mu^a$,
$\tilde{ \bar{s}} A_\mu^a$, $\tilde s c^a$, $\tilde{\bar{s}}\bar c^a$
and $\tilde s \bar c^a$ denote arbitrary operators depending on
$\{A,c,\bar c,h,\alpha,\beta,\bar \beta\}$, of dimension two, with the
same transformations under linear symmetries as the corresponding bare
expressions. In order for $\tilde s$ and $\tilde{\bar s}$ to be the
symmetries of $\hat \Gamma$ discussed just below equation (\ref{ST}),
we complement their definitions (again by analogy with
(\ref{lagsources}) and (\ref{symmbrst})) by
\begin{align}
\label{eqsh}\tilde s\,i h^a&=\frac 12 \frac{\delta \hat \Gamma}{\delta \bar \beta^a} \\
\tilde {\bar s}\,i h^a&=-\frac 12 \frac{\delta \hat \Gamma}{\delta\beta^a}\\
\tilde s \bar c^a-\tilde  {\bar s} c^a&=2i h^a.
\end{align}
For generic operators, one defines $\tilde s$ as
\begin{equation}
\label{tildes}
\begin{aligned}
\tilde s&=\int d^4x\Big\lbrace
\tilde sA_\mu^a(x) \frac{\delta}{\delta A_\mu^a(x)}
+\tilde sc^a(x) \frac{\delta}{\delta c^a(x)} \\
&\hspace{.5cm}+\tilde s\bar c^a(x) \frac{\delta}{\delta \bar c^a(x)}
+\tilde s h^a(x)\frac{\delta}{\delta h^a(x)}
\Big\rbrace
\end{aligned} 
\end{equation}
and similarly for $\tilde{\bar s}$. It is now easy to check that, with
these definitions, $\tilde s$ and $\tilde{\bar s}$ are symmetries of
$\hat \Gamma$.
 
We now want to solve the Slavnov-Taylor equation (\ref{ST}) together
with Eqs. (\ref{eqmoth},\ref{eqmotc},\ref{eqmotA}). The calculation is
lengthy but straightforward. Some details are given in the
Appendix. The resolution simplifies if one introduces the variables
\begin{align}
\label{tildes2}
\tilde c^a&=c^a+Z_L \beta^a \nonumber\\
\tilde {\bar c}^a&=\bar c^a+Z_L \bar\beta^a \nonumber\\
\tilde A_\mu^a&= A_\mu^a-Z_K \alpha_\mu^a \\
i\tilde h^a&=ih^a+\frac {Z_L}2\Big((\tilde D_\mu\alpha_\mu)^a+\tilde g
f^{abc}(\bar c^b \beta^c-\bar \beta^b c^c)\Big),\nonumber
\end{align}
where $(\tilde D_\mu \phi)^a=Z\partial_\mu\phi^a+\tilde g
f^{abc}\tilde A_\mu^b \phi^c$. $Z$ and $\tilde g$ are at this level
arbitrary constants. In term of these variables, the solution reads:
\begin{widetext}
\begin{equation}
\label{eqgammahat}
\begin{split}
\hat \Gamma=&\int d^4x\Bigg\{\frac{\tilde Z}4\tilde F_{\mu \nu}^a\tilde F_{\mu \nu}^a + \frac
{Z_L}{2Y} (\partial_\mu \tilde A^a_\mu)^2+\frac
{(\tilde D_\mu \tilde{\bar c})^a \partial_\mu
\tilde c^a + \partial_\mu \tilde{\bar c}^a (\tilde D_\mu
\tilde c)^a }{2 Y} -  \frac{\tilde g^2 \left(f^{abc}\tilde{\bar c}^b\tilde c^c\right)^2}{8 Y}\\
&+\frac {\xi_0}2 \tilde h^a \tilde h^a +i\tilde h^a \partial_\mu \tilde
A_\mu^a+\bar\beta^a\tilde gf^{abc}\left(i\tilde h^b \tilde c^c+\frac{\tilde
  g}4 f^{cde}  \tilde{\bar c}^b\tilde c^d\tilde c^e\right)
+\tilde gf^{abc}\left(i\tilde h^b \tilde {\bar c}^c-\frac{\tilde
  g}4 f^{cde}  \tilde{ c}^b\tilde {\bar c}^d\tilde { \bar
  c}^e\right)\beta^a
\\&- \frac{Z_L}4\left((\tilde
D_\mu\alpha_\mu)^a-\tilde gf^{abc}(\bar \beta^b \tilde c^c-\tilde
{\bar c}^b \beta^c)\right)^2-\alpha_\mu^a\left(i(\tilde D_\mu \tilde
h)^a+\frac{\tilde g}2 f^{abc}(\tilde{\bar c}^b (\tilde D_\mu \tilde
c)^c -(\tilde D_\mu \tilde{\bar c})^b  \tilde c^c)\right)
\Bigg\},
\end{split}
\end{equation}
\end{widetext}
with 
$\tilde F_{\mu\nu}^a=Z (\partial_\mu \tilde A_\nu^a- \partial_\nu
\tilde A_\mu^a)+\tilde g f^{abc}\tilde A_\mu^b \tilde A_\nu^c$ and $Y=1-Z_L \xi_0/2$.
The action of $\tilde s$ and $\tilde {\bar s} $ on the fields reads:
\begin{align}
\label{previous2}
\tilde s A_\mu^a &= (\tilde D_\mu \tilde c)^a , \nonumber\\
\tilde{ \bar{ s}} A_\mu^a &=(\tilde D_\mu \tilde{\bar c})^a, \nonumber\\
\tilde s c^a &= -\frac{{\tilde g}}{2}f^{abc} \tilde c^b \tilde c^c, \nonumber\\
\tilde{\bar{s}}\bar c^a &= -\frac{{\tilde g}}{2} f^{abc} \tilde{\bar c}^b \tilde{\bar c}^c, \nonumber\\
\tilde {\bar{s}} c^a\,&=-ih^a-\frac{{\tilde g}}{2}f^{abc} \tilde{\bar c}^b \tilde c^c,\nonumber\\
\tilde s \bar c^a&=ih^a-\frac{{\tilde g}}{2}f^{abc} \tilde{\bar c}^b \tilde c^c.
\end{align}
Note that equation (\ref{eqgammahat}) is written in terms of the bare
gauge parameter $\xi_0$. The reason being that Eq. (\ref{eqmoth})
ensures that the $h$-sector of the effective action is not
renormalized. Actually, Eqs. (\ref{eqmoth},\ref{eqmotc},\ref{eqmotA})
impose two other relations:
\begin{align}
\label{gtilde}
 g_0&= \tilde g Y, \nonumber \\
 1+Z_K&= Z Y.
\end{align}
These equations are at the core of the non-renormalization theorem
(see below).

The action of $\tilde s$ on $h$ can be deduced from
Eq. (\ref{eqsh}). We just give here the expression at vanishing
sources for the composite operators:
\begin{equation}
\begin{split}
2 iY \tilde s h^a&=i{\tilde g}Y^2f^{abc}h^bc^c-\xi_0
       \frac{Z_L^2}4\tilde g f^{abc}\partial_\mu A_\mu ^b c^c
\\&+\frac{\tilde
  g^2}{4 }f^{abc}f^{cde} \bar c^b c^dc^e -
       Z_L(\tilde D_\mu\partial_\mu c)^a
\end{split}
\end{equation}
An analogous formula can be derived for $\tilde{\bar s}h$. 

A straightforward calculation shows that $\tilde s$ and $\tilde{\bar
  s}$ are nilpotent on-shell, {\it i.e.} when one imposes the
equations of motion for the fields $h$, $c$ and $\bar c$. Actually
$\tilde s$ and $\tilde{\bar s}$ can be decomposed in a sum of an
off-shell nilpotent symmetry that has the form of the bare symmetry
(\ref{symmbrst}) up to multiplicative factors and two trivial symmetries with
generators:
\begin{align}
r_1c=&r_1\bar c=r_1A_\mu=0 \nonumber\\
r_1h^a=&-f^{abc }\frac{\delta \hat\Gamma}{\delta h^b}c^c,
\end{align}
and
\begin{align}
r_2A_\mu&=r_2c=0 \nonumber\\
r_2\bar c^a&=-i\frac{\delta  \hat\Gamma}{\delta h^a} \nonumber\\
r_2\,i h^a&=-\frac{\delta  \hat\Gamma}{\delta \bar c^a}.
\end{align}
These generators vanish when one imposes the equations of motion. This
is consistent with the on-shell nilpotency of $\tilde s$ and $\tilde{\bar s}$.

Note that there appears in $\Gamma$ terms that where not present in
the bare action described in Section~\ref{chap_model}. There are terms
with two powers of the sources or more and also a term proportional to
$(\partial_\mu A_\mu^a)^2$.  In order to make the theory
renormalizable, one needs to include such terms in the bare
action. Fortunately, it is not necessary to perform again the analysis
with this new action. Indeed, the precise form of the bare action is
not necessary to deduce Slavnov-Taylor identities. All what is needed
is that the bare action satisfies the Slavnov-Taylor identities
\cite{Zinn-Justin}. Therefore, the form of $\Gamma$ given in
Eqs. (\ref{previous1},\ref{eqgammahat},\ref{previous2}) is stable
under renormalization. Let us comment that the term in $(\partial_\mu
A_\mu^a)^2$ can be eliminated by a shift proportional to $\partial_\mu
A_\mu^a$ of the Lagrange multiplier.

We now make contact with the perturbative results and concentrate on
the $A$, $c$, $\bar c$ sector once the Lagrange multiplier has been
eliminated by its equation of motion. The standard parametrization
(see for instance \cite{Gracey02}) of the effective action reads:
\begin{align}
\hat \Gamma&=\int d^4x \Bigg\{ \frac{1}{2 Z_c}\left( \partial_\mu \bar
c^a\breve D_\mu c^a+
 \breve D_\mu \bar c^a\partial_\mu c^a\right)\nonumber\\
&+\frac{1}{2\xi_0 Z_\xi}(\partial_\mu A_\mu^a)^2
-\frac{Z_\xi \xi_0 g_0^2}{8 Z_g^2 Z_A Z_c^2}( f^{abc}\bar c^b c^c)^2 \nonumber\\
&+ \frac{1}{4 Z_A} \breve F^a_{\mu\nu}\breve F^a_{\mu\nu}\Bigg\}.
\end{align}
with 
\begin{align}
\breve D_\mu c^a&=\partial_\mu c^a + \frac{g_0}{Z_g \sqrt{Z_A}} f^{abc} A_\mu^b c^c, \nonumber\\
\breve D_\mu \bar c^a&=\partial_\mu \bar c^a + \frac{g_0}{Z_g \sqrt{Z_A}} f^{abc} A_\mu^b \bar c^c, \nonumber\\
\breve F^a_{\mu\nu}&=\partial_\mu A_\nu^a-\partial_\nu A_\mu^a+\frac{g_0}{Z_g \sqrt{Z_A}} f^{abc} A_\mu^bA_\nu^b.
\end{align}
Comparison with Eq. (\ref{eqgammahat}) -- where $h$ is eliminated by
its equations of motion -- yields, together with Eq. (\ref{gtilde}),
the following relations:
\begin{align}
\label{our-world}
Z_A&=Z^{-2}\tilde Z^{-1} \nonumber\\
Z_c&=Y Z^{-1}\nonumber\\
Z_\xi&=Y\nonumber\\
Z_g&=YZ^2\tilde Z^{1/2}.
\end{align}
One then easily deduces the non-renormalization theorem:
\begin{equation}
\label{th_renorm_1}
Z_g= Z_A^{-1/2}Z_c^{-1} Z_\xi^{2}.
\end{equation}
We postpone to section \ref{perttheory} the discussion of this
equation together with another non-renormalization theorem to be
proven in the next section.

\section{The massive case}
\label{chap_mass}
As said in the introduction, Curci and Ferrari proposed a very natural
generalization of Yang-Mills theory in this particular gauge
\cite{Curci76,Curci76b}. One can add a mass term for the ghosts and
gluons that preserves BRST-like symmetries:
\begin {equation}
\label{lagmass}
\mathcal L_{m}=m_0^2\left(\frac 12 (A_\mu^a)^2+\xi_0 \bar c^a
c^a\right).
\end {equation}
The theory remains renormalizable; however, nilpotency of the BRST
symmetry is lost and, as a result, the model is no longer unitary
\cite{Curci76b,Ojima81,deBoer95}. The study performed in the previous
sections is generalized here to include the mass term
(\ref{lagmass}). We show that the modifications to
Eqs. (\ref{eqmoth},\ref{eqmotc},\ref{eqmotA}) are very simple. The
other striking result is that no independent renormalization factor is
needed to renormalize the mass term.

Let us start by discussing the symmetry content of the theory in the
presence of the mass term. All the linear symmetries (a--d) are
preserved. On the contrary the action is not invariant under the
original BRST and anti-BRST transformations (\ref{symmbrst}), but are
invariant under modified transformations, $s_{m}=s+m_0^2 \,s_{1}$ and
$\bar s_{m}=\bar s+m_0^2\,\bar s_{1}$, with
\begin{align}
&s_{1}c^a=s_{1}\bar c^a=s_{1}A_\mu^a=0 \nonumber \\
&\bar s_{1}c^a=\bar s_{1}\bar c^a=\bar s_{1}A_\mu^a=0 \nonumber \\
&s_{1}\, ih^a=c^a \nonumber \\
&\bar s_{1}\, ih^a=\bar c^a.
\end{align}
As already mentionned the new BRST and anti-BRST symmetries
transformations are no longer nilpotent. Its algebra becomes
\cite{Delduc89}
\begin{align}
s_{m}^2&=m_0^2 \,t \nonumber \\
\bar s_{m}^2&=m_0^2 \,\bar t \nonumber \\
\{s_{m},\bar s_{m}\}&=-m_0^2\, N.
\end{align}

The Curci-Ferrari mass term induces a change in the Slavnov-Taylor
equation. The right-hand side of (\ref{ST}) is not zero anymore and
must be replaced by a term proportional to $m_0^2$:
\begin{equation}
\begin{split}
m_0^2\int d^4x\Bigg( i&\frac{\delta \Gamma}{\delta h^a} c^a+
\alpha_\mu^a\frac{\delta \Gamma}{\delta\bar K_\mu^a}\ \\&-
2\bar\beta^a\frac{\delta \Gamma}{\delta \bar L^a}+ 2 \frac{\delta
  \Gamma}{\delta M^a}\beta^a\Bigg).
\end{split}
\end{equation}
Slight modifications must be introduced to the Ward identities
described in Section \ref{chap_model}. The equation (\ref{eqmoth}) is
actually not modified because the mass term (\ref{lagmass}) is
independent of $h$. In the second identity, Eq. (\ref{eqmotc}), one
must add $-\xi_0 m_0^2 c^a$ to the left-hand side. Finally, in
Eq. (\ref{eqmotA}), one must add $- m_0^2 \partial_\mu A_\mu^a$ to the
left-hand side.

One easily checks that the divergent part (\ref{previous1}) of the
solution of the modified Slavnov-Taylor equation now reads
$\Gamma_{m,div}^{(n)} =\Gamma_{div}^{(n)}+m_0^2\,\Gamma_{1,div}^{(n)}$
with
\begin{equation}
\begin{split}
\Gamma_{1,div}^{(n)}=\int& d^4x\Bigg(\frac {\frac 12(\tilde
A_\mu^a)^2+Z\xi_0 \bar{\tilde
        c}^a\tilde c^a}{Z Y}\\&-\frac
      {Z_K}2(\tilde\alpha_\mu^a)^2 +2Z_L\bar{\tilde\beta}^a\tilde\beta^a\Bigg).
\end{split}
\end{equation}
Note that the renormalization of the mass term does not require any
new renormalization factor. This leads to another non-renormalization
theorem. If one compares the previous equation at zero sources with
the standard parametrization of the mass term \cite{Gracey02},
\begin{equation}
\int d^4x\frac {m_0^2}{Z_m}\Bigg(\frac {
A_\mu^aA_\mu^a}{2Z_A }+\frac{\xi_0 Z_\xi}{Z_A Z_c} \bar{
        c}^a c^a\Bigg),
\end{equation}
by identification of the $A^2$ terms one deduces that
\begin{equation}
 Z_m Z_A= Z Y.
\end{equation}
The $\bar c c$ term does not give new information. Using the
identifications (\ref{our-world}), one obtains another
non-renormalization theorem:
\begin{equation}
\label{th_renorm_2}
Z_\xi^2=Z_m Z_A Z_c.
\end{equation}

\section{Consequences for perturbation theory}
\label{perttheory}

The two non-renormalization theorems presented in previous sections
have far reaching consequences for practical perturbative
calculations. First of all, they imply that one has to calculate as
many renormalization factors in CFDJ gauge as in linear
gauges. Moreover, all these renormalization factors can be extracted
from the 2-point function of gluons alone. In fact, one possible set
of independent renormalization factors are $Z_m$ (the renormalization
for the composite operator $A_\mu^aA_\mu^a$), $Z_A$ and $Z_\xi$ that
can all be extracted from the zero momentum, transverse and
longitudinal parts at order $p^2$ of the quoted correlation function.
Other choices may even be more convenient in practice since some of
these renormalization factors can be extracted from the 2-ghost
function that has simpler kinematics. In any case, there is no need to
calculate 3-point or higher vertices, contrarily to what is required
in linear gauges. The price to pay is very small: there is a 4-ghost
vertex, but the required total number of diagrams seems to be always
smaller than that in linear gauges. For example, the 1-loop beta
function for pure gauge, can be extracted from three diagrams
only. So, in what concerns perturbative calculations, once
non-renormalization theorems are exploited, CFDJ gauge is as
competitive as linear gauges (same number of renormalization factors)
and might even be more convenient (all renormalization factors can be
extracted from 2-point functions).

To conclude, let us make three final remarks. First, these two
renormalization factors have been found previously by one of
us. However the proof of these non-renormalization theorems presented
in \cite{Wschebor07} requires extensive use of equations of motions
because it is formulated without the introduction of the Lagrange
multiplier field $h$. The physical content of these identities is
therefore hidden. Here, these relations are shown to be
consequences of the new Ward identities. Second, one can check that
the 3-loop renormalization factors \cite{Gracey02} satisfy the two
non-renormalization theorems (\ref{th_renorm_1},
\ref{th_renorm_2}). Actually, it was observed in \cite{Browne06} that the 3-loops
renormalization factors satisfy the identity (\ref{th_renorm_2}) without
giving a general proof.
Finally, for the Landau gauge ($\xi=0$), $Z_\xi=1$
and one recovers the well-known non-renormalization theorem for the
coupling constant \cite{Taylor71} as well as the more recent one for
the mass \cite{Dudal02}.

\section{Superspace interpretation}
\label{superspace}

\subsection{Flat superspace}
It has been shown in the 80's that reinterpreting the theory in a
superspace enables one to give a geometric meaning to the symmetries
of the model, in particular to BRST and anti-BRST symmetries
\cite{Bonora80, Bonora81, Baulieu81, Delbourgo81}. We review here the
superfield formalism and subsequently reinterpret the new Ward
identities described in the previous section in this context.

In the following, we consider a $4+2$ dimensional superspace, with $4$
standard bosonic coordinates, noted $x^\mu$, and 2 grassmanian --
anticommuting -- coordinates $\ts$ and $\tb$:
$\ts^2=\tb^2=\ts\tb+\tb\ts=0$.  The (super)fields are now functions of
$x^\mu$, $\ts$ and $\tb$. In the following, capital indices vary on the
4 bosonic directions $\mu$ and on the 2 grassmanian directions: for
instance $x^M=(x^\mu,\ts,\tb)$. Because of the grassmanian character
of $\ts$ and $\tb$, the Taylor-expansion in powers of these variables
gives a finite number of terms:
\begin{equation}
\begin{aligned}
f(x^\mu,\ts,\tb)=&f_{00}(x^\mu)+\ts f_{10}(x^\mu)+\tb
f_{01}(x^\mu)\\&+\tb\ts f_{11}(x^\mu),
\end{aligned}
\end{equation}
with
$f_{ij}(x^\mu)=\partial_\ts^i\partial_\tb^jf(x^\mu,\ts,\tb)|_{\ts=\tb=0}$.
Observe that the derivatives with respect to $\ts$ and $\tb$ are
nilpotent, just as BRST and anti-BRST symmetries. It is actually
possible to make this analogy stronger, if one writes a $4$-dimensional
field $\phi$ and its BRST/anti-BRST variations as a $4+2$ dimensional
superfield $\Phi$
\begin{equation}
\begin{aligned}
\Phi(x^M)=&\phi(x^\mu)+\tb\ s\phi(x^\mu)-\ts\ \bar
s\phi(x^\mu) \\&+\tb \ts \ s \bar s\phi(x^\mu),
\end{aligned}
\end{equation}
where now it is clear that $s$ and $\bar s$ act on the superfield as
$\partial_\tb$ and $-\partial_\ts$ respectively.

Note moreover that the vectorial superfields, like the gauge field,
have $4+2$ components $\mathcal A^\mu$, $\mathcal A^\ts$ and $\mathcal
A^\tb$, which have ghost numbers 0, 1 and -1 respectively. One can
therefore merge the 4-dimensional gauge field, the ghost, anti-ghost
and all BRST/anti-BRST variations of these fields in a unique vectorial
superfield:
\begin{align}
\label{orthproj}
\mathcal A^\mu(x^M)=&A^\mu+\tb sA^\mu-\ts \bar
sA^\mu +\tb \ts\  s \bar sA^\mu \nonumber\\
\mathcal A^\ts(x^M)=&c+\tb\ sc-\ts\ \bar
sc +\tb \ts \ s \bar sc \nonumber\\
\mathcal A^\tb(x^M)=&\bar c+\tb\ s\cb-\ts\ \bar
s\cb +\tb \ts \ s \bar s\cb,
\end{align}
where we have omitted the color index and the bosonic space
variable. The BRST/anti-BRST symmetries can therefore be interpreted
as the invariance under translation in the grassmanian directions. It
is important to understand at this level that the components $\ts$,
$\tb$ and $\ts\tb$ of the fields $\mathcal A^\mu$, $\mathcal A^\ts$
and $\mathcal A^\tb$ are not independent of the $\ts=\tb=0$ part of
the fields. Indeed, these are explicit functions of $A^\mu$, $c$,
$\bar c$ and $h$, as given in
Eqs.~(\ref{symmbrst},\ref{variationssb}). Consequently, the superfield
is constrained and cannot be used as it stands in a functional
integral. These constraints are sometimes called ``transversality
conditions'' \cite{ThierryMieg79,Bonora80, Bonora81, Baulieu81,
  Delbourgo81}.

The symmetries $t$ and $\bar t$ and $N$ given in Eq.~(\ref{symmsp2})
also have a simple geometric interpretation in superspace: they
correspond to the invariance under ``rotations'' in the grassmanian
directions.

The Lagrangian is easily recast in terms of superspace and
superfield. One finds for instance \cite{Delbourgo81}
\begin{equation}
\label{slaggf}
\mathcal L_{\rm{GF}}=-\int d\ts d\tb\ \frac 12  \mathcal A^Mg_{MN}\mathcal A^N,
\end{equation}
with $g$ a metric in the superspace, defined as
\begin{equation}
\label{metricflat}
g_{MN}=
\begin{cases}
\delta_{\mu\nu}&\rm{if\ } M=\mu,\ N=\nu\\
-\xi_0/2&\rm{if\ } M=\ts,\ N=\tb\\
\xi_0/2&\rm{if\ } M=\tb,\ N=\ts\\
0&\rm{otherwise.}
\end{cases}
\end{equation}
The gauge-fixing term appears formally as a mass term in the
theory. Observe that $\xi_0$ appears as a different normalization of
the bosonic and fermionic coordinates that can be reabsorbed by a
change of variables, in the same way as the speed of light can be
eliminated in Minkowskian space.

The source term can also be written as \cite{Delbourgo81}
\begin{equation}
\label{slagsources}
\mathcal L_{\rm{sources}}=\int d\ts d\tb\  \mathcal
A^Mg_{MN}\mathcal J^N
\end{equation}
with
\begin{align}
  \mathcal J^\mu&=\alpha^\mu-\tb K^\mu-\bar K^\mu\ts+\tb\ts J^\mu\nonumber\\
  \frac {\xi_0} 2\mathcal J^\ts&=\beta+\ts \frac{M-iR}2+\tb L+\tb\ts \chi\nonumber\\
  \frac {\xi_0} 2\mathcal J^\tb&=\bar \beta-\tb \frac{M+iR}2-\ts \bar
  L+\tb\ts \bar \chi.
\end{align}

The Yang-Mills term does not have such a nice superspace
expression. One can however write it as \cite{Delbourgo81}
\begin{equation}
\label{slagym}
\mathcal L_{\rm{YM}}=\int d\ts d\tb\  \frac 14 \tb \ts (\mathcal F_{\mu\nu}^a)^2
\end{equation}
with 
\begin{equation}
\mathcal F_{\mu\nu}^a=\partial_\mu\mathcal
A_\nu^a-\partial_\nu\mathcal A^a_\mu +g_0 f^{abc}\mathcal A^b_\mu \mathcal
A^c_\nu.
\end{equation}

After this review of the supersace formalism, let us now come to the
interpretation of the supergauge symmetries described in Section
\ref{chap_model}. The infinitesimal gauge transformations can actually
be written in the very concise form:
\begin{equation}
\delta \mathcal A_M^a=\partial_M \Lambda^a+g_0 f^{abc}\mathcal A_M^b
\Lambda ^c,
\end{equation}
where $\Lambda$ is an arbitrary function of $x^M$. This transformation
has exactly the same form as a standard gauge transformation in
Yang-Mills theory. To make contact with the expressions of Section
\ref{chap_model}, we just need to write the Taylor expansion of
$\Lambda$ in powers of $\ts$ and $\tb$:
\begin{equation}
\label{supergaugeparam}
 \Lambda(x^M)=\lambda(x^\mu)+\tb\eta(x^\mu)+\bar
 \eta(x^\mu)\ts+\tb\ts\hat \lambda(x^\mu).
\end{equation}

This transformation not only gives the right gauge transformation of
the physical fields $A$, $c$, $\bar c$ and $h$, but also give
consistent gauge variations for their BRST and anti-BRST
variations. Moreover, the Ward identities have a very natural
interpretation. Indeed, the Yang-Mills part of the action
(\ref{slagym}) is manifestly invariant under the supergauge
transformation. The gauge-fixing term (\ref{slaggf}) breaks this
symmetry. However its variation under (\ref{supergaugeparam}) is
linear in the field,
\begin{equation}
\delta \mathcal L_{\rm{GF }}=-\int d\ts d\tb\ \mathcal
A^{a,M}\partial_M\Lambda^a,
\end{equation}
and one can therefore deal with it in the corresponding Ward identity.

\subsection{Curved superspace}

If the superspace formulation of the massless CFDJ model has been
known for quite some time, the corresponding formulation for the
massive CF model has never been addressed before. This is the aim of
this subsection. The important observation in this respect is that the
BRST and anti-BRST transformations $s_{m}$ and $\bar s_{m}$ (that were
associated with translations in the grassmanian sector in the massless
case) do not anticommute. Their anticommutator is indeed proportional
to $m_0^2$ times a ``rotation'' in the grassmanian coordinates. This
is very similar to what happens when one studies the commutation
relations of the rotations of the sphere in the limit of infinite
radius, where the sphere approaches a plane. At leading order in the
curvature, two rotations can be interpreted as translations (that
commutes) and the third corresponds to the rotation of the plane. We
therefore expect that the theory in the presence of a mass term is
associated with a superfield theory in a curved superspace, with
curvature proportional to $m_0$.

The calculations in a curved superspace requires the introduction of a
formalism similar to the one of General Relativity. Actually all the
standard formulas in a curved space have their superspace equivalent
that differ by some signs. We followed the formalism and conventions
of \cite{Nath75,Arnowitt75}, except that we work with
left-derivatives. In particular, we consider the supercovariant
derivative of a superverctor $\mathcal{V}^N$:
\begin{equation}
 \mathcal{D}_M \mathcal{V}^N=\partial_M \mathcal{V}^N+\Gamma^N_{MP} \mathcal{V}^P
\end{equation}
with Christoffel symbols
\begin{align}
\label{chris}
\Gamma^C_{AB}=&\frac{(-1)^{bc}}{2}\Big((-1)^{ab+b}\partial_B g_{AD}+(-1)^{b}\partial_A g_{BD}\nonumber\\
&-(-1)^{d(a+b)+d}\partial_D g_{AB}
 \Big) g^{DC}.
\end{align}
Here and below, the lowercase letters are 1 if the associated
uppercase is fermionic and 0 otherwise.  The covariant derivative
$\mathcal D_M$ should not be confused with the derivative $D_\mu$
associated with the gauge group, which we used up to now.

As in standard Riemann geometry, superspace symmetries are described
by the Killing vectors that satisfy the equation
\begin{equation}
 \mathcal{D}_M \mathcal{X}_N+(-1)^{mn} \mathcal{D}_N \mathcal{X}_M=0.
\end{equation}
Taking the Lie bracket of two Killing vectors $\mathcal{X}$ and $\mathcal{Y}$,
\begin{equation}
 [\mathcal{X},\mathcal{Y}]^M=\mathcal{X}^P \partial_P \mathcal{Y}^M-\mathcal{Y}^P \partial_P \mathcal{X}^M
\end{equation}
gives another Killing vector. The corresponding algebra is the Lie
algebra of the isometry group of the superspace. Moreover, the Killing
vectors generate the infinitesimal field transformations under
isometries again by the Lie bracket:
\begin{equation}
 \mathcal{A}^M\to \mathcal{A}^M+\epsilon[\mathcal{X},\mathcal{A}]^M.
\end{equation}

In the following we consider the metric
\begin{equation}
g_{MN}=
\begin{cases}
\delta_{\mu\nu}&\rm{if\ } M=\mu,\ N=\nu\\
-\frac{\xi_0}{2}(1+m_0^2\tb\ts)&\rm{if\ } M=\ts,\ N=\tb\\
\frac{\xi_0}{2}(1+m_0^2\tb\ts)&\rm{if\ } M=\tb,\ N=\ts\\
0&\rm{otherwise.}
\end{cases}
\end{equation}
Observe first that it identifies with (\ref{metricflat}) in the limit
$m_0\to 0$. Moreover, it is compatible with Poincar\'e and symplectic
symmetry groups but does not respect the translation invariance in
grassmanian coordinates. From Eq. (\ref{chris}) one can deduce that
the non-zero Christoffel symbols are:
\begin{align}
 \Gamma^{\ts}_{\ts\tb}&=-\Gamma^{\ts}_{\tb\ts}=-m_0^2 \,\ts \nonumber\\
 \Gamma^{\tb}_{\ts\tb}&=-\Gamma^{\tb}_{\tb\ts}=-m_0^2 \,\tb.
\end{align}
Using the expression for the scalar curvature of ref.~\cite{Nath75},
one finds that the superspace has a finite and homogeneous scalar
curvature $R=-12 m_0^2/\xi$.

In order to verify that it does correspond to the CF model we
calculated the most general Killing vector, obtaining:
\begin{align}
\mathcal{X}^\mu&=a^\mu+R^{\mu\nu}x^\nu,\nonumber\\
 \mathcal{X}^\theta&=\alpha(1+m_0^2\bar\theta\theta)+\bar\theta \beta-\theta\delta,\nonumber\\
 \mathcal{X}^{\bar \theta}&=\bar \alpha(1+m_0^2\bar\theta\theta)+\bar\beta \theta +\bar \theta\delta.
\end{align}
The part proportional to $a^\mu$ corresponds to translations and the
one proportional to $R^{\mu\nu}=-R^{\nu\mu}$ to rotations in bosonic
coordinates. The parts proportional to $\bar \beta$ and $ \beta$
correspond to the symmetries $t$ and $\bar t$ respectively, while the
part proportional to $\delta$ corresponds to the ghost
number. Finally, the parts proportional to $\bar \alpha$ and $\alpha$
correspond to BRST and anti-BRST symmetries respectively (observe that
they become translations when $m_0\to 0$).  By a straightforward
calculation one can verify that the Lie bracket of the Killing vectors
generate the Lie algebra of symmetries of the CF model as described in
section \ref{chap_mass}. It is also an easy task to verify that the
Killing vectors generate the right fields transformations for the
fields $A, c, \bar c$ and $h$ as defined in section
\ref{chap_mass}. Finally, one can verify that the only renormalizable
lagrangian compatible with the symmetries of the curved superspace is
that of the CF model.

\section{Conclusion}
\label{conclusion}

In the present paper, we have shown that the CFDJ gauge fixing of
Yang-Mills theory verifies four non trivial local and linear Ward
identities. This result has many consequences. First, it allows the
deduction of two non-renormalization theorems, that reduces the number
of independent renormalization factors from five to
three. Consequently, in perturbation theory, one has to calculate as
many renormalization factors as in linear gauges. Moreover, as
discussed in section \ref{perttheory}, all these renormalization
factors can be extracted from the 2-point functions alone. We expect
that this simplifies considerably the perturbative calculations in
Yang-Mills theory.

Another important result of the present paper is that the obtained
Ward identities can be interpreted in the superfield formalism for
Yang-Mills theory as consequences of supergauge transformations. The
generalization to the theory with a CF mass term is simple and it is
shown to be equivalent in the superfield formalism to a curvature of
the superspace in the grassmannian coordinates. Up to now, however,
the superfields are constrained by the so-called ``transversality
condition''. As a consequence, one cannot use them as they stand in a
functional integral. Let us stress that the existence of this
supergauge symmetry reinforces our conviction that the superfield
formalism is of prime importance in this field. This pushes one to look
for the description of the model in terms of unconstrained
superfields. This work is currently in progress.

\begin{acknowledgments}
We thank B. Delamotte, M. Reisenberger and G. Tarjus for useful
discussions. M. T. thank the IFFI for hospitality, where most of this
work was done. We acknowledge support of PEDECIBA and the ECOS
program. N. W. acknowledge support of PDT uruguayan program.  LPTMC is
UMR 7600 of CNRS
\end{acknowledgments}

\appendix
\section{Solving Slavnov-Taylor identity}

In this appendix we give some details of the derivation of equations
(\ref{tildes}) and (\ref{eqgammahat}). We first substitute the
expression (\ref{previous1}) into the Slavnov-Taylor identity
(\ref{ST}) and analyze the terms quadratic in the sources $K,L,\bar K,
\bar L$ and $M$. One easily finds that $\tilde s A$ and $\tilde s c$
do not depend arbitrarily on $c$, $A$, $\beta$ and $\alpha$, but
only through $\tilde c^a$ and $\tilde A_\mu^a$ (see Eq. (\ref{tildes2})).

If one now study the terms linear in $K,L,\bar K, \bar L$ and $M$, one
finds four independent constraints. The two relations
\begin{align}
\label{nil1}
&\tilde s^2 A_\mu^a=0, \nonumber \\ &\tilde s^2 c^a=0,
\end{align}
give the nilpotency in a particular sector. 
One also finds
\begin{equation}
\label{eqssba}
\begin{split}
\tilde s \tilde{\bar s} A_\mu^a(x)=&-Z_K \frac{\delta \hat \Gamma}{\delta A_\mu^a(x)}-\frac{\delta \hat \Gamma}{\delta \alpha_\mu^a(x)} \\
&-\frac{Z_L}{2} \Big(\tilde D_\mu \frac{\delta \hat \Gamma}{\delta h}\Big)^a.
\end{split}
\end{equation}
Adding to this equations its conjugate one deduces:
\begin{equation}
\{\tilde s,\tilde{\bar s}\}A_\mu^a=0,
\end{equation}
which again expresses the nilpotency in another sector. The fourth
relation reads
\begin{equation}
\label{eqssbcb}
\tilde s \tilde {\bar s}\bar c^a=-Z_L\frac{\delta \hat \Gamma}{\delta
  c^a}+\frac{\delta \hat \Gamma}{\delta \beta^a}-i\frac {{\tilde g}Z_L}2
\frac{\delta \hat \Gamma}{\delta h^b}f^{abc} \tilde {\bar c}^c.
\end{equation}

Now, the most general operators of dimension two, respecting Lorentz
invariance, global color invariance, the symmetry (\ref{symmsp2}),
ghost number conservation, the definition (\ref{eqsh}) and nilpotency
(\ref{nil1}) are written in (\ref{previous2}).

Equations (\ref{eqssba},\ref{eqssbcb}) take a simpler form if one
introduces the variable $\tilde h^a$ defined in Eq.~(\ref{tildes2}).
Taking as independent variables $\tilde A_\mu$, $\tilde c$, $\tilde
{\bar c}$, $\tilde h$, $\alpha_\mu$, $\beta$ and $\bar\beta$ one deduces
\begin{align}
\label{sbarsAc}
\tilde s \tilde {\bar s}\bar c^a&=\frac{\delta \hat \Gamma}{\delta \beta^a} \nonumber \\
\tilde s \tilde {\bar s}  A_\mu^a&=-\frac{\delta \hat \Gamma}{\delta \alpha_\mu^a}.
\end{align}
Note that at this level we have explicit expressions for the $\tilde
s$ and $\tilde{\bar s}$ variations of fields $A, c$ and $\bar c$ but
not $h$. However, the left hand sides of equations (\ref{sbarsAc}) can
be computed without knowledge of the variations of $h$. Therefore, we
have an explicit expression for the derivatives of $\hat \Gamma$ with
respect to $\alpha$ and $\beta$ (and by conjugation of $\bar
\beta$). One can then integrate these trivial differential equations
and obtain the dependence of $\hat \Gamma$ on these variables. As a
result, we only need to find the part of $\hat \Gamma$ that does not
depend on the sources. The dependence on $h$ (and then on $\tilde h$)
is trivially deduced from (\ref{eqmoth}). The remaining part is
obtained by imposing the invariance of $\hat \Gamma$ under $\tilde
s$. One finally obtains the result (\ref{eqgammahat}).

\bibliographystyle{unsrt}

\end{document}